\documentclass[conference]{IEEEtran}
\IEEEoverridecommandlockouts
\usepackage{setspace}
\usepackage{xcolor}
\usepackage[flushleft]{threeparttable}
\usepackage{floatrow}
\usepackage{url}

\usepackage{graphicx}
\usepackage{amssymb}
\usepackage{epstopdf}
\usepackage{physics}
\usepackage{algpseudocode}
\usepackage{algorithm}
\usepackage{cite}
\usepackage{enumitem}

\usepackage{url}
\usepackage{amsfonts,amssymb,amsmath}
\usepackage{bm,theorem,color}

\usepackage{array}
\usepackage{calc}
\usepackage{cases} 
\usepackage{multirow}
\usepackage{tcolorbox}
\usepackage{epstopdf}
\usepackage{tabularx}
\usepackage{tabulary}
\usepackage{amsmath}
\usepackage{multirow}
\usepackage{commath}
\usepackage{subfig}

\usepackage{xcolor}

\usepackage{mathtools}
\usepackage{newtxmath}

\usepackage{csquotes}

\begin{document}

\title{Integrating Hardware Security into a Blockchain-Based Transactive Energy Platform 
\thanks{This study was funded by the United States Office of Naval Research (ONR) Defense University Research-to-Adoption (DURA) Initiative under award number N00014-18-1-2393.}
}

\author{\IEEEauthorblockN{Shammya Shananda Saha\IEEEauthorrefmark{1}\IEEEauthorrefmark{3}, Christopher Gorog\IEEEauthorrefmark{5}, Adam Moser\IEEEauthorrefmark{5}, Anna Scaglione\IEEEauthorrefmark{1},
Nathan G. Johnson\IEEEauthorrefmark{2}}
\IEEEauthorblockA{\IEEEauthorrefmark{1}School of Electrical, Computer, and Energy Engineering, Arizona State University
\\\IEEEauthorrefmark{5}BlockFrame Inc. 
\\\IEEEauthorrefmark{2}The Polytechnic School, Arizona State University\\
Email: \IEEEauthorrefmark{3}shammya.saha@asu.edu}
}
\maketitle

\begin{abstract}
This applied research paper introduces a novel framework for integrating hardware security and blockchain functionality with grid-edge devices to establish a distributed cyber-security mechanism that verifies the provenance of messages to and from the devices. Expanding the idea of Two Factor Authentication and Hardware Root of Trust, this work describes the development of a Cryptographic Trust Center\textsuperscript{TM} (CTC\textsuperscript{TM}) chip integrated into grid-edge devices to create uniform cryptographic key management. Product managers, energy system designers, and security architects can utilize this modular framework as a unified approach to manage distributed devices of various vendors, vintages, and sizes. Results demonstrate the application of CTC\textsuperscript{TM} to a blockchain-based Transactive Energy (TE) platform for provisioning of cryptographic keys and improved uniformity of the operational network and data management. This process of configuring, installing, and maintaining keys is described as Eco-Secure Provisioning\textsuperscript{TM} (ESP\textsuperscript{TM}). Laboratory test results show the approach can resolve several cyber-security gaps in common blockchain frameworks such as Hyperledger Fabric.
\end{abstract}


\section{Introduction}
The potential of the Smart Grid rests on advanced information and communication technologies to enable and enhance electric grid efficiency, reliability, and resilience. An enabler of this future is having security built-in (rather than bolted onto) the associated hardware and network protocols. This motivation is increasing given the proliferation of grid-edge devices such as distributed energy resources (DERs), electric vehicles (EV), smart loads, and grid-level energy storage. These devices form part of the emerging Internet-of-Things ecosystem that enables Transactive Energy (TE) in which energy trading occurs at the grid-edge. This approach allows local economic signals to elicit a response that increases grid efficiency and removes congestion within the electrical network.

Blockchain is an ideal approach to enable a transactive network because grid-edge devices come from a broad matrix of vendors and types, with units that can be added or removed from the network at any time, use rule sets that may be updated frequently, and lack a centralized authority for management and control. It allows non-trusting market participants to trust each through utilizing a common immutable transaction record validated by several peers. It offers a flexible, low cost, and secure means to implement logistics and tracking architecture to manage digital assets and distributed devices \cite{gorog_solving_2018}. This feature addresses one of the most critical issues tied to the security of interconnecting grid-edge devices with critical electricity infrastructure. Prior works have shown the use of a permissionless blockchain (e.g., Ethererum) \cite{munsing_optimization,Jonathan2018,FOTI2019113604} and a permissioned blockchain (e.g., Hyperledger Fabric or HLF) \cite{Wang2019,Kim2020,Karandikar2019} for TE. This paper addresses the following unresolved security issues in HLF-based TE architectures:
\begin{enumerate}[leftmargin=*]
\item HLF uses the JSON Web Token (JWT) authentication method based on claim token \cite{jwt}. However, as BASE64 is used as the encoding method without data encryption in claim tokens, malicious users can collect sensitive information by eavesdropping on access tokens of genuine users \cite{Park_2018}.
\item HLF provides security credentials for prosumers, but the grid-edge devices owned by prosumers can still be subject to attacks such as physical attacks, communications disruptions, and firmware reprogramming. An excellent example is discussed in \cite{spectrum2015} in which a smart inverter vendor remotely updated control settings for 800,000 inverters in a single day in Hawaii. Were an adversary able to penetrate the back-end system in the same fashion, or hijack the update, that attacker could manipulate power flows or disable or damage equipment, and perhaps cause a localized or grid-wide blackout.
\end{enumerate}
This paper addresses these concerns by integrating hardware security via a Cryptographic Trust Center\textsuperscript{TM} (CTC\textsuperscript{TM}) chip. Design considerations for such a hardware chip are provided in \cite{Saleem2019DesignCO} to protect grid-edge devices; however, the solution is expensive and prevents mass scaling. Expanding the idea of Two Factor Authentication (2FA) \cite{Park_2018, Eman_2019}, and Hardware Root of Trust \cite{Casper2011}, the proposed solution integrates hardware security in conjunction with the existing security features of a blockchain-based TE. The approach is demonstrated for a physical solar photovoltaic (PV) inverter, an Energy Management System (EMS) at the prosumer location, and a HLF-based blockchain network managing the TE market in the cyber domain.

\section{Blockchain-Based Transactive Energy Platform}
\label{ref:blockchain_te}
\begin{figure}
    \centering
    \includegraphics[width=0.9\columnwidth]{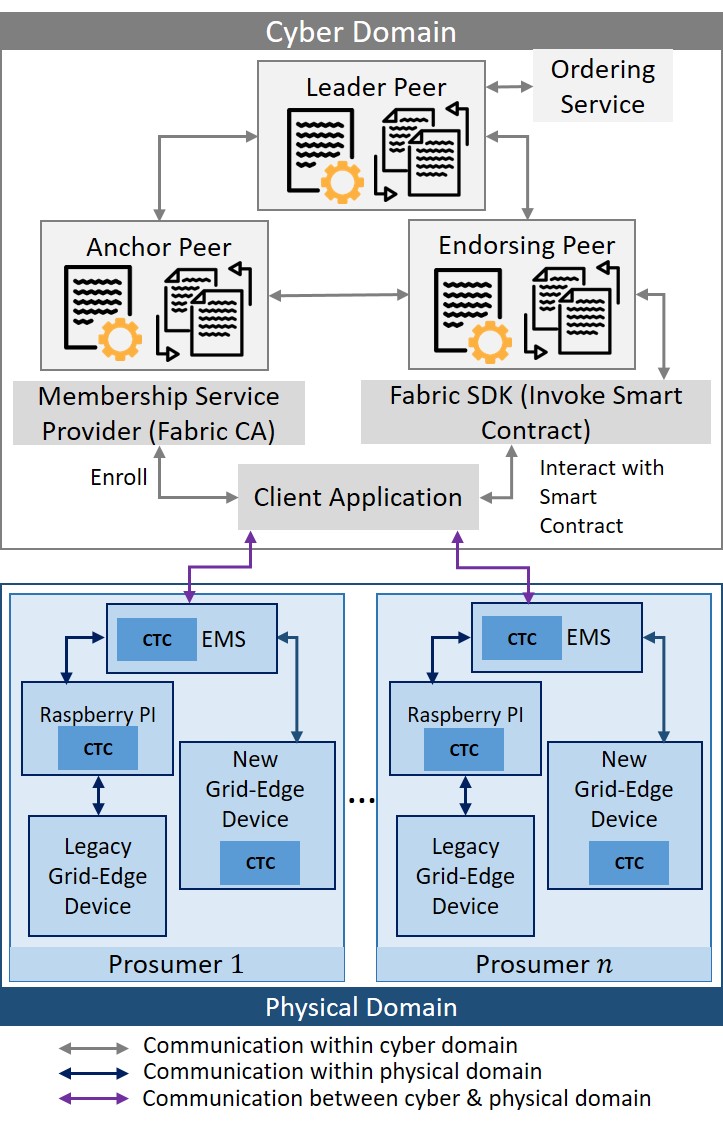}
    \caption{A Cyber Physical Architecture for Blockchain Based Transactive Energy Network}
    \label{fig:cyber_physical}
\end{figure}

Figure \ref{fig:cyber_physical} shows a cyber-physical architecture where the cyber domain is built using HLF, and the physical domain consists of market participants or prosumers who own various types of grid-edge devices. Each prosumer is a \textquote{client} to the blockchain network that leverages client applications to interact with the network. 

\begin{figure}
    \centering
    \includegraphics[width=0.6\columnwidth]{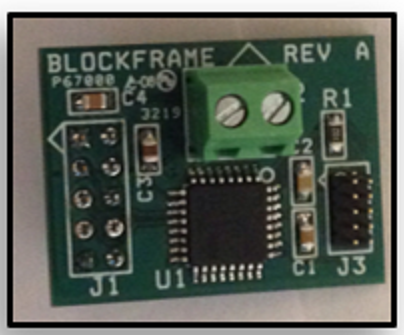}
    \caption{Cryptographic Trust Center\textsuperscript{TM} Chip}
    \label{fig:bfchip}
\end{figure}

We assume that each prosumer is equipped with an EMS responsible for managing the underlying physical assets in the physical domain. The EMS is equipped with the CTC\textsuperscript{TM} (Figure \ref{fig:bfchip}). Physical assets may include legacy assets or new (smart-enabled) infrastructure. Legacy assets integrate with the framework via a Raspberry Pi equipped with the CTC\textsuperscript{TM} chip installed post-manufacture. New grid-edge devices can have the CTC\textsuperscript{TM} installed directly by the manufacturer, thereby skipping the Raspberry Pi for marshaling data from the grid-edge device to the EMS.  

The approach provides a secure method to distribute cryptographic keys to grid-edge devices. We call this approach to configuring, installing, and maintaining keys as Eco-Secure Provisioning\textsuperscript{TM} (ESP\textsuperscript{TM}). Importantly, we incorporate blockchain technologies to record \textit{each provisioning event} and maintain immutable records of such transactions for all grid-edge devices. The logistics and tracking architecture for managing the physical domain through ESP\textsuperscript{TM} is handled through a blockchain that is \textit{separate} from the blockchain managing the cyber domain.

In this work, we have developed a HLF-based TE solution that is capable of performing the basic functionalities: 
\begin{enumerate} [leftmargin=*]
    \item Collecting bids from market participants/prosumers;
    \item Handling an iterative distributed pricing algorithm;
    \item Allowing the local EMS to collect dispatch set points from the TE network through smart contracts and passing those set points onto underlying physical asset(s);
    \item Maintaining a record of all transactions for billing.
\end{enumerate}
The first function requires interaction between the cyber and physical domain, and the third function requires interactions within the physical domain. The remaining two functions remain within the cyber domain and hence require no interaction with the physical domain. Any interactions between the cyber and physical domains are considered as a \textquote{transaction}. This work shows how the CTC\textsuperscript{TM} chip and ED25519 verification \cite{Khovratobich_2017} expands upon 2FA to integrate hardware security into those transactions. 

\section{Security Features Using Eco-Secure Provisioning\textsuperscript{TM} for Cryptographic Key Management}
\label{sec:bpf}
\begin{figure*}
    \centering
    \includegraphics[width = \textwidth]{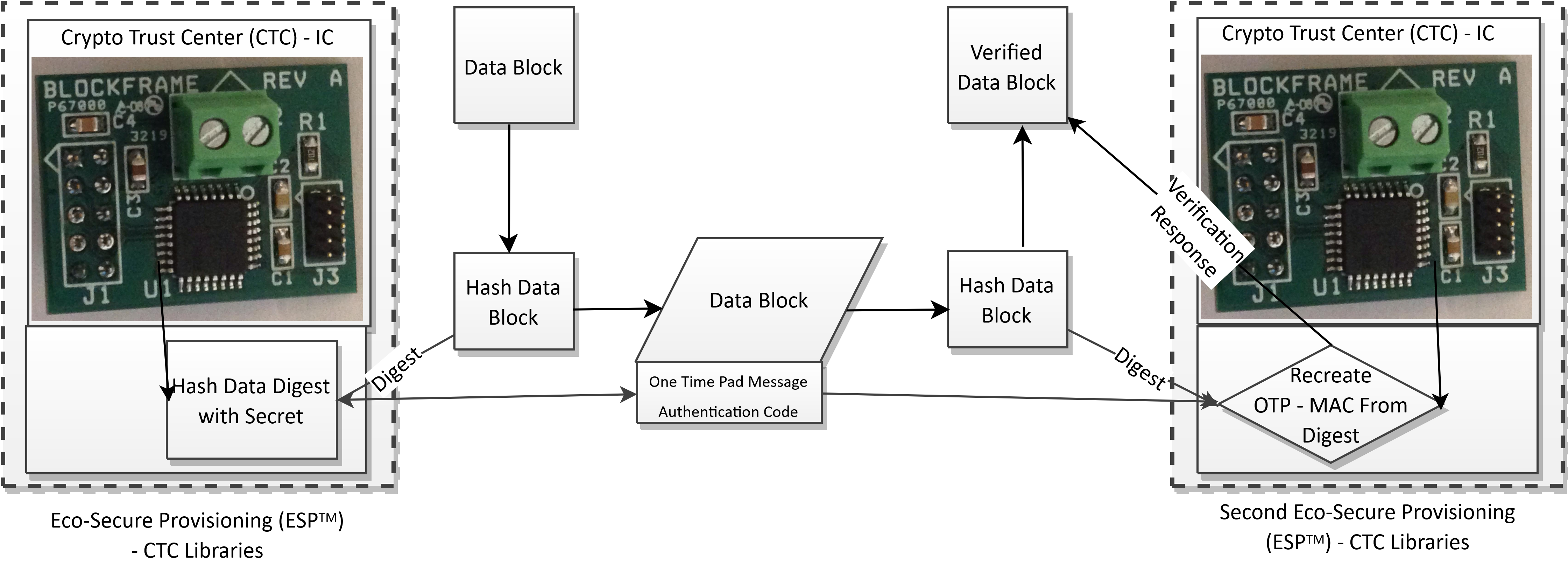}
    \caption{OTP Message Authentication}
    \label{fig:otp_message}
\end{figure*}
The ESP\textsuperscript{TM} framework includes a small integrated circuit placed within each grid-edge device. A cloud-based Industry Utility Registrar\textsuperscript{TM} (IUR\textsuperscript{TM}) service offers \texttt{Trust as a Service} to provision trusted components within each device \cite{hoor_trust_2011}. The provisioning process ensures that no human can access provisioned components, thus preventing hostile actors from circumventing system security without triggering global awareness.

\subsection{Cyber-Security Layers}
The integrity of the cyber-security approach is enabled by the ESP\textsuperscript{TM} framework that uses provisioning protected by logistical tracking of keys contained in each distributed grid-edge device. A cryptographically-secure distributed ledger maintains the record of cryptographic keys providing visibility of provisioned instances while protecting the actual keys and private values and verifying content without compromising itself. The framework supports any application needs for asynchronous and synchronous communications and supports each case with asymmetric and symmetric crypto-logic operations. 
\begin{itemize}[leftmargin=*]
\item \textbf{Application Layer Secure} – Virtual Private Networks (VPN) sessions are used between any two provisioned devices to give application-level secure communications for one-to-one systems or organizational virtual local area network (VLAN) connections. 
\item \textbf{Data Proof of Origin} – A protected signing capability is used to create Proof of Origin (POO) signatures for any data item. POO forensically proves the device's identity that created it and can do so at any time throughout the life-cycle of the data. 
\item \textbf{Timed Challenge-Response} – Time-restricted challenge-response enforced sequences provide the means to assure that unsanctioned computational analyses on exchanged command components are less likely during the sequence. This is valuable for situations with strict time restrictions when devices perform sensitive or dangerous operations, such as electro-mechanical actuation commands. 
\item \textbf{Uniform Network Segments} – Ensuring operational availability requires network segments such as a local area network, to operate uniformly in the event of the loss of wide-area communications. The framework’s distributed uniformity helps sustain the persistence of secure operation without reliance on continuous communication to key management services.
\end{itemize}
\begin{figure*}
        \centering
        \includegraphics[width=0.9\textwidth]{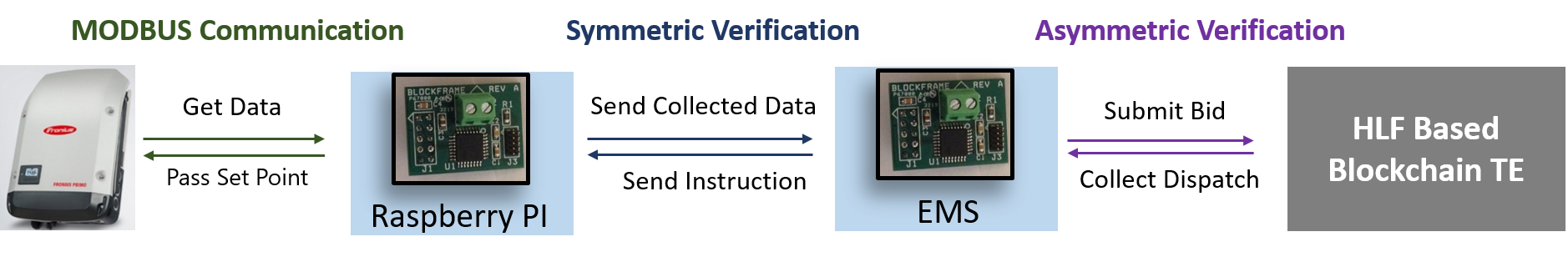}
        \caption{Architecture of the Hardware Integration Test}
        \label{fig:architecture}
\end{figure*}
\subsection{Organization-wide Security Support}
With a new approach referred to as individualized-uniformity, the framework provisions each device uniquely in a uniform method such that they all work together. Each device can independently perform the sets of security controls. However, other uniform components allow for a synergistic operation of all the devices acting together as a system of systems to increase overall organization-wide security effectiveness. Whether designed for inclusion at the factory or integrated onto legacy applications, the ESP\textsuperscript{TM} framework can support the following functions.
\begin{itemize}[leftmargin=*]
\item \textbf{Zero-Trust Networking} – The support for individualized-uniformity enables Zero-Trust Networking capabilities because all remote devices can reject unapproved devices, users, or software versions by default until that device can self-validate with multi-factor identification \cite{flanigan_zero_2018,kindervag_build_2010}. 
\item \textbf{Trusted Certificate Provisioning} – Operations that use Transport Layer Security (TLS) connections are strengthened by the distribution and provisioning of trusted certificates \cite{barker_recommendations_2018,nist_glossary_2013}. ESP\textsuperscript{TM} framework provisions valid certificates and adds features to assure the accuracy of Certificate Revocation Lists (CRL), making legacy TLS security features more robust.
\item \textbf{Patch and Update Verification} – Individualized confidentiality controls, enabled by software signatures, can be created for restricted use by individual devices. Single-use codes and restricted-content control can be used to assure that software updates are only installed after verified as trusted.
\end{itemize}

\section{Security Features in Grid-edge Devices with Cryptographic Trust Center\textsuperscript{TM} Chips}
\label{sec:2fa}
Integration of ESP\textsuperscript{TM} capabilities with HLF based blockchain TE for transmission of data was accomplished using both asymmetric and symmetric capabilities as detailed below: 
\begin{itemize}[leftmargin = *]
\item \textbf{Asymmetric} – These operations use public key cryptography where a mathematically related \textquote{public key} and \textquote{private key} are used. Only the private key needs to remain secret. This is a common approach of digital signature algorithms in which the private key and data are used to generating a unique signature. The recipient can then independently verify the integrity of the underlying signed data using the sender's public key.
\item \textbf{Symmetric} – These operations rely on both parties possessing the same cryptographic key, which no outside party can access. For example, most encryption standards use symmetric algorithms that require less system resources, thus enabling faster operations (compared to the asymmetric case) and near run-time performance. Symmetric verification also provides support for checking the authenticity of underlying devices.
\end{itemize}
\subsection{Single-Use One Time Pad Verification – Symmetric Message Authentication}
In symmetric message authentication, transmitted data is sent with One Time Pad (OTP) Message Authentication Code (MAC) to ensure sensitive command messages remain intact. Upon receiving a message and MAC from a sender, a recipient device equipped with a CTC\textsuperscript{TM} chip can verify the sender's MAC. MAC messages can be sent alongside any command or response to continuously verify either the receiving or the sending device or in the middle of any transmission sequence. Security in this approach requires both the IUR\textsuperscript{TM} and device to possess the same cryptographic key, which is used to create and verify the MAC. Since the symmetric cryptographic key never leaves the CTC\textsuperscript{TM} chip (cryptographic operations are performed inside), no attacker can clone or impersonate the device, if they gain access to the device’s memory or hard drive for a while. Devices which possess the same symmetric key are also able to MAC with each other directly.

The base algorithm used for symmetric authentication is SHA3-512 \cite{sha}, a hash function that, when combined with a secret key, can be used both as a message authenticator and Key Derivation Function. The sequence of operation for symmetric verification is shown in Figure \ref{fig:otp_message}. 

\subsection{Signature Creation and Verification – Digital Signing for Data Integrity Verification}
Information from a grid-edge device used for data analytic, controls, user billing transactions, and more functions requires high levels of reliability and integrity. The provisioning framework provides each enabled device a unique elliptic curve asymmetric key pair (private and public key). The asymmetric algorithm used on the chip is ED25519, an elliptic curve with a security level comparable to 3072-bit Rivest–Shamir–Adleman (RSA) cryptosystem. Each enabled device signs the data to be exchanged using the private key through an elliptic curve digital signature algorithm. Thus, data from each device can be verified at collection points or after periods of storage using the device's public key. Private keys are used for signing, but do not leave the CTC\textsuperscript{TM}. After signing, data is transmitted to the blockchain network managing the TE. Smart contracts can verify the integrity of the received information using the public key through ED25519 verification. Hence, the public key belonging to CTC\textsuperscript{TM} is tied into the blockchain’s credential management system without compromising the private keys which remain in the chips. The developed solution is independent of the blockchain framework used to run the TE market because the solution designed does not explicitly depend on the TE's underlying blockchain architecture. As long as ED25519 verification is possible from the blockchain smart contract, the CTC\textsuperscript{TM} can be used to provide additional verification similar to 2FA verification. A detailed example of a market transaction that uses asymmetrical verification is explained in Section \ref{sec:asymmetric}. 

\section{Hardware Integration and Demonstration}
\label{sec:hardware}
The proposed framework was demonstrated using Hardware-In-the-Loop (HIL) simulation described next. 
\subsection{Experimental Setup}
\label{sec:setup}
The HIL simulation includes an EMS that interacts with TE blockchain to submit consumption and generation bids and to collect dispatch points. The EMS communicates with a legacy inverter through the Raspberry Pi and MODBUS protocol. A $4$ KW Fronius Primo inverter was used during the demonstration. Code for managing the EMS is written in \texttt{Python} using Django web framework \cite{django2020}. Though this implementation illustrates secured communication with one physical asset, the Django web framework is suitable for incorporating additional physical assets to the EMS. The Raspberry Pi is equipped with the hardware CTC\textsuperscript{TM} chip and uses I$^\textnormal{2}$C protocol to communicate with the chip. Figure \ref{fig:architecture} shows the communication links between the associated components with appropriate verification algorithms described in Section \ref{sec:2fa}. 

\begin{figure}
    \centering    
    \includegraphics[scale=0.8]{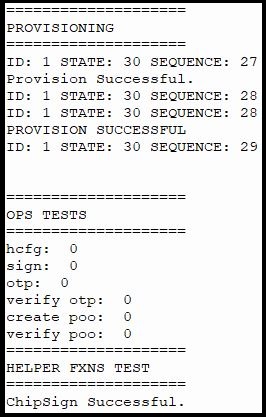}
    \caption{Provisioning of the CTC\textsuperscript{TM} Chip}
    \label{fig:provisioning}
\end{figure}

\subsection{Provisioning of the CTC\textsuperscript{TM} Chip}
\label{sec:initialization}
The provisioning process implements a sequence for updating the CTC\textsuperscript{TM} from the supply chain provisioned status to the end-user application provisioned status. The provisioning IUR\textsuperscript{TM} and CTC\textsuperscript{TM} implement this process using a protocol and sequence consisting of three distinct message-segments, as illustrated in Figure \ref{fig:provisioning}.

In the first step, both the IUR\textsuperscript{TM} and local CTC\textsuperscript{TM} produce random inputs creating symmetric mutual authenticated challenge-response. This process utilizes shared secrets provisioned within CTC\textsuperscript{TM} based on default values, which would be entered into the CTC\textsuperscript{TM} by a manufacturing vendor. These shared secrets are also maintained within the IUR\textsuperscript{TM} server. 

The second message-segment contains the packet of cryptographic key information destined for the CTC\textsuperscript{TM}, which is encrypted and can only be interpreted by the CTC\textsuperscript{TM}. Upon receiving the second message, the CTC\textsuperscript{TM} inserts the new cryptographic configuration in temporary status, awaiting the third step. 

The third message-segment completes a cryptographic key super-session \cite{nist_super_session} sequence by verifying both old and new cryptographic configurations are available within the CTC\textsuperscript{TM} by using them together. Upon receiving the super-session verification, the IUR\textsuperscript{TM} creates a final commit stream and returns an encapsulated stream containing this sequence. The CTC\textsuperscript{TM} finalizes super-session and commits the new cryptographic configuration for use only if all signatures for the final sequence content are verified successfully. 
\begin{figure*}[]
    \centering
    \includegraphics[width=1\textwidth]{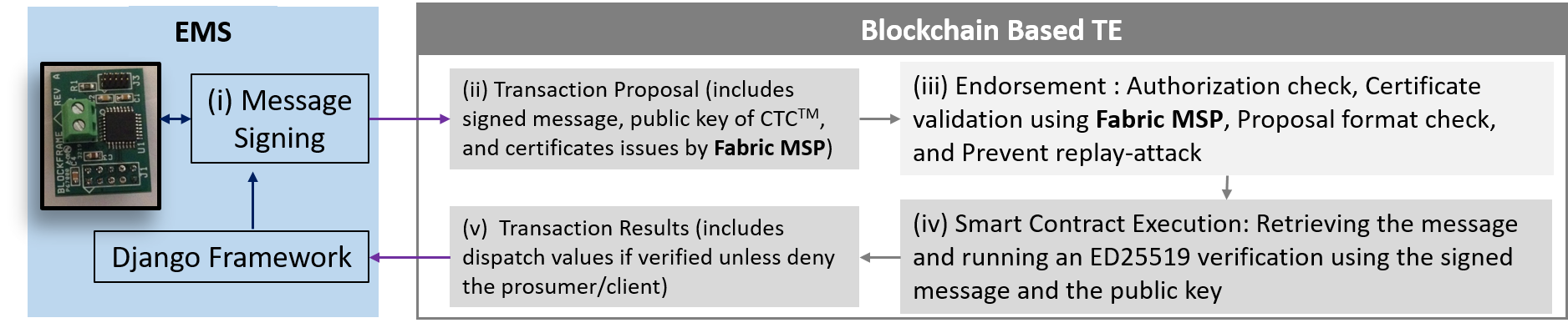}
    \caption{Two Factor Authentication Using Asymmetric Verification}
    \label{fig:ed25519}
\end{figure*}

\begin{figure}
    
	\centering
	\subfloat[]{%
	\label{fig:incoming}%
	\includegraphics[width=\columnwidth]{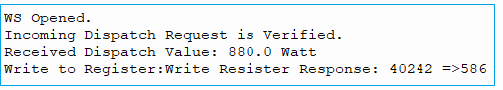}
	}%
	\\
	\subfloat[]{%
	\label{fig:fronious}%
	\includegraphics[scale=0.17]{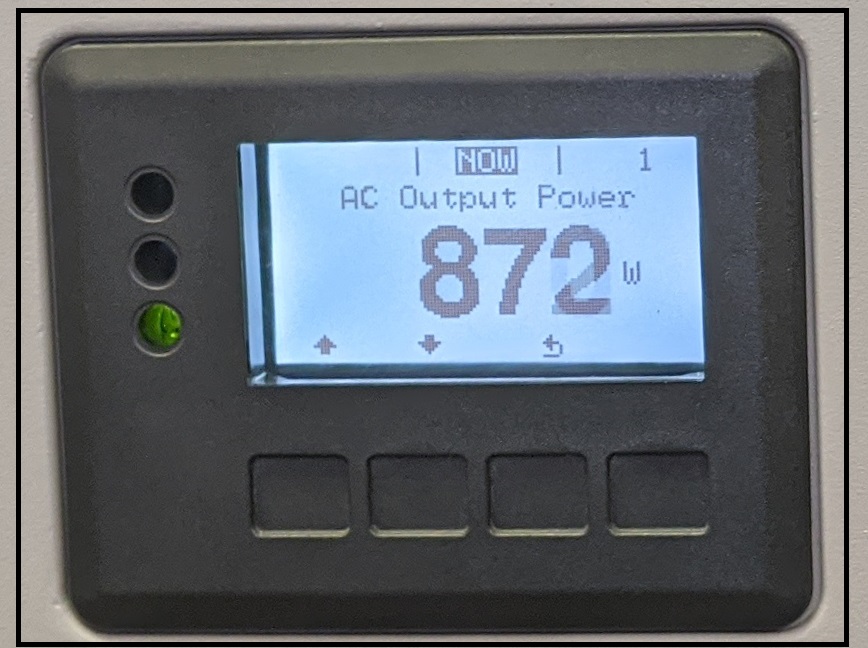}
	}%
	\\
	\subfloat[]{%
	\label{fig:controller}%
	\includegraphics[width=1\columnwidth,keepaspectratio]{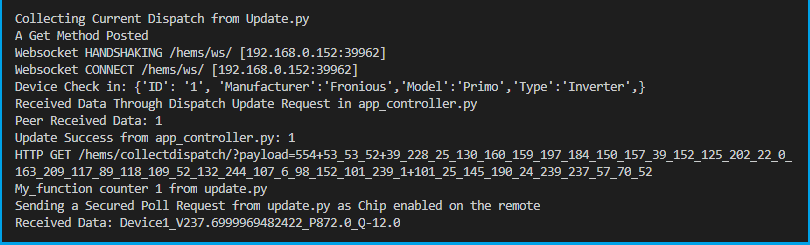}
	}%
	\caption{\protect\subref{fig:incoming} EMS Sends the Dispatch Value to the Raspberry Pi \protect\subref{fig:fronious} Output from the Fronius Inverter \protect\subref{fig:controller} EMS Receives Information after Symmetric Verification}
	\label{fig:symmetric_model}
\end{figure}
\subsection{Asymmetric Verification in TE Network Communications}
\label{sec:asymmetric}
An example use case is demonstrated here for asymmetric verification that allows integrating hardware security with the blockchain smart contract. Figure \ref{fig:ed25519} illustrates the implementation while the EMS is collecting the dispatch schedule from the TE blockchain. 

The EMS is assigned a unique ID with credentials issued by a Hyperledger Fabric Certificate Authority. The EMS uses the private key to sign the message where the message includes the prosumers ID. Using the \texttt{Node.js} client SDK, a transaction proposal is created that contains the signed message and the chip's public key as the input arguments. After submitting the transaction, the endorsement process in HLF verifies the client's (i.e., EMS's) credential using the Membership Service Provider. If verified, the smart contract then takes input arguments from the transactions and uses the public key to perform an ED25519 verification that verifies the prosumer's ID against the credentials issued by the Certificate Authority. This second verification is analogous to the idea of 2FA, which provides additional security to secure the TE network against malicious prosumers. A similar procedure is followed when the prosumer submits a bid that includes the prosumer's unique ID, type (DER/ electric vehicle/ controlled load, and more), and associated parameters. 

\subsection{Symmetric Verification for Two-way Asset Communications}
\label{sec:symmetric}
After collecting the dispatch set point, the EMS sends this information to the legacy inverter through the Raspberry Pi and the Django framework. Sending data through the Django framework requires the use of \texttt{websocket} \cite{websocket} protocol, which can transmit a JSON formatted data object. Hence the EMS creates a JSON object that includes the message \texttt{type} (to query information from the inverter or to send dispatch instruction) and the message itself. For example, while sending the dispatch information, the message includes the real power dispatch value and the OTP created by the EMS. Upon receiving, the Raspberry Pi parses the message and uses the underlying instruction, and it's own cryptographic key to generate it's OTP. If the generated OTP matches the OTP embedded in the message, the Pi accepts the command and responds accordingly. The process is repeated on the EMS after receiving the response from the Raspberry Pi. This symmetrical verification secures communication between the EMS and any physical asset communicating with the EMS. Figure \ref{fig:symmetric_model} shows the use of OTP while sending dispatch information from the EMS to the inverter and receiving output from the inverter\footnote{The difference between the command sent and the output of the inverter happens due to rounding issue in the inverter and intermittence in solar irradiance.}.

\section{Discussion}
This paper introduced a process for cryptographic key provisioning and demonstrated laboratory evaluation of a hardware chip that embeds cyber-security into grid-edge devices to enable secure peer-to-peer transactive energy trading. The approach and chip were shown to integrate with legacy assets, and could also be installed by manufacturers on new assets, to allow a rapid vendor-agnostic approach to scale cyber-security worldwide while providing governance support on an industry-wide scale. The hardware implementation also integrated a blockchain to address security gaps using two-factor identification and hardware root of trust. Although transactive energy was the motivating case demonstrated in the paper, the approach can be used to secure any data transmitted such as firmware updates, energy use readings, price signals, and more. This leaves significant flexibility in integrating this solution with other services that need to verify and trust grid-edge devices that can be added or removed from the network.

\bibliographystyle{IEEEtran}
\bibliography{conference_041818.bib}

\end{document}